\documentclass{article}
\usepackage[utf8]{inputenc}
\usepackage{lmodern}
\usepackage{graphicx}
\usepackage{caption}
\usepackage{booktabs}
\usepackage{subfig}
\usepackage{hyperref}
\usepackage{rotating}
\usepackage{pdflscape} 
\usepackage{array} 

\DeclareCaptionLabelFormat{boldfig}{\textbf{Fig #2}}
\title{Improving the accuracy of food security predictions by integrating conflict data}
\author{Marco Bertetti, Paolo Agnolucci, Alvaro Calzadilla, Licia Capra}
\date{}

\begin{document}

\maketitle

\begin{abstract}
Violence and armed conflicts have emerged as prominent factors driving food crises. However, the extent of their impact remains largely unexplored. This paper provides an in-depth analysis of the impact of violent conflicts on food security in Africa. We performed a comprehensive correlation analysis using data from the Famine Early Warning Systems Network (FEWSNET) and the Armed Conflict Location Event Data (ACLED). Our results show that using conflict data to train machine learning models leads to a 1.5\% increase in accuracy compared to models that do not incorporate conflict-related information. The key contribution of this study is the quantitative analysis of the impact of conflicts on food security predictions.

\end{abstract}

\section{Introduction}
Food security (FS) is a complex and multifaceted problem, influenced by several factors such as weather events, economic shocks, and natural disasters. Understanding the dynamics of food security is crucial for effective policymaking and humanitarian efforts. While conflicts and violent events increasingly stand out as key drivers of food crises\cite{WFP-GlobalReport2024}, the depth of their impact remains largely underexplored. Examining the quantitative aspects of this impact is essential for developing more targeted interventions and strategies to address the complex interplay between conflict and food security. Existing research tends to be qualitative in nature (Kemmerling et al.2022; Brown et al. 2020; Brown et al. 2021), leaving a significant gap in understanding the quantitative aspects of how conflicts impact FS levels. By delving into quantitative analyses, we can not only enhance our comprehension of the magnitude of the problem but also pave the way for evidence-based decision-making in efforts to alleviate food insecurity in conflict-affected 
regions. \\

Regarding the qualitative study of conflicts and FS, Kemmerling et al.(2022)\cite{kemmerling2022} provided a comprehensive explanation on how violence and armed conflicts impact FS through destruction, displacement, financing of conflicts and food being used as a weapon. The authors call for better conflict data collection, and an increase in focus on the study of conflicts early warnings. This work offers a broad and insightful view on how conflict data can be better utilised and included into FS predictions. The authors suggest that more and better integration of conflict data may make FS predictions more accurate. \\

Brown et al. (2020)\cite{Brown2020} and Brown et al. (2021)\cite{brown2021} highlighted that research on the effect of conflicts on child health is lacking, further confirming the lack of coverage for this topic in the broader FS literature. They also pointed out that the long-term effect of conflicts on malnutrition is still not well understood, and how data regarding conflicts need improvement in both collection and adoption in early warning systems.\\
These studies contribute valuable insights into the complex interplay between armed conflicts and food security (FS). However, they do not quantify the overall impact of conflicts on food security, nor demonstrate how conflict data can be effectively used to enhance food security predictions and modeling. \\

In a study commissioned by the FAO, Brück et al. (2016)\cite{bruck2016} investigated the intricate relationship between conflicts and various proxy indicators of food security (FS). The study focused on two case studies in Ethiopia and Somalia, providing empirical evidence of the reciprocal connection between food security and conflict dynamics. This work highlights the critical issue of bidirectionality, demonstrating that conflicts can exacerbate food insecurity, while food insecurity can also contribute to the onset or escalation of conflicts. Although the direction of causality may vary depending on the context, evidence from developing countries in Africa predominantly indicates that conflicts precede and aggravate food security challenges. This aligns with the causal direction assumed in our research. In Ethiopia, the onset of violence at the local level is associated with a decline in agricultural production, whereas in Somalia, conflict is correlated with wasting, as indicated by anthropometric indicators. While these empirical studies mark a significant advancement compared to purely qualitative analyses, it is important to note that this work does not actively contribute to the enhancement of food security predictions.\\

Andreé et al. (2020)\cite{Andree2020} are amongst the first scholars to begin integrating conflict data, such as the count of violent events and their intensity (fatalities) into machine learning (ML) models to make predictions on the transition between non-crisis and crisis state of FS. Data regarding violent events was combined with heterogeneous data, such as population and land data, and Normalized Difference Vegetation Index (NDVI), a remote sensing measurement commonly used to assess vegetation health and density using satellite imagery. However, this work does not provide insights into the specific value added by integrating conflict data when making predictions, as conflict data in their model represents only a small subset of the variables used, which include several heterogeneous datasets. Consequently, the effect of conflict data on FS predictions is not isolated, making it unclear how much this particular data source contributes to the overall model performance.\\

This study aims to fill these gaps by providing the first quantitative assessment of the impact of conflict on FS across multiple geographic levels. To achieve this, we employ a novel dataset that integrates conflict data from the Armed Conflict Location Event Data Project (ACLED) with food security assessments from the Famine Early Warning Systems Network (FEWSNET). By combining these data sources, we introduce a comprehensive, data-driven framework for examining the relationship between conflict events and FS outcomes.

We advance the field methodologically by employing multi-level correlation analysis and machine learning models to quantify the impact of conflict data on FS predictions. This approach enables a more granular understanding of how conflict dynamics influence food security at country, regional, and district levels, providing targeted insights that were previously unattainable with existing methods. This approach aligns with the pressing need for a data-driven understanding to inform policy decisions, enhance early warning systems, and develop effective predictive models for mitigating food crises.\\

This paper is structured as follows: the next section offers a brief qualitative overview, highlighting the key mechanisms through which conflicts influence food security. This overview sets the foundation for the quantitative analysis by illustrating the various ways in which conflicts can disrupt food security across different contexts. The subsequent section details the applied methods for the quantitative study, encompassing the approach adopted for both correlation analysis and modeling work. The metrics employed throughout the study are then described in the subsequent section. Moving forward, the results section shows the empirical findings, providing a tangible illustration of the study's outcomes. The concluding section offers closing remarks, summarizing key insights and implications drawn from the investigation.

\section{The Broad Impact of Conflicts on FS}
Wars and conflicts, whether at a national or localized level, significantly impact food security by disrupting agricultural production, food distribution, and accessibility. According to the World Food Programme (WFP) \cite{wfp2020}, "Conflict is the single biggest driver of hunger in the world today...UN Resolution \#2417 recognizes that humankind will never eliminate hunger without establishing peace in the world."\footnote{Hunger, conflict, and improving the prospects for Peace, WFP, October 2020}.

Conflicts can reduce overall food availability by disrupting agricultural production and supply chains, as exemplified by the 2022 Russian invasion of Ukraine. This conflict affected global food security through the destruction of agricultural infrastructure, restrictions on exports, and disruptions in the availability of agricultural inputs, such as fertilizers \cite{fao2022, csis2022}. 
Moreover, conflicts lead to surging food prices, making food less affordable, particularly in developing regions where a large proportion of disposable income is spent on basic sustenance. As seen during the ongoing Ukraine conflict, the FAO Food Price Index reported record increases in global food prices \cite{fao2022_Prices}.

Localized conflicts can also directly disrupt communities' ability to produce or access food. Displacement caused by violence leaves agricultural lands unproductive, as in the central Sahel region, where millions have been forced to abandon their livelihoods \cite{Raleigh2020, eu2022}. Additionally, deliberate blockades, such as those in the Tigray region of Ethiopia, prevent the delivery of food aid, further exacerbating hunger \cite{economist2021}. \\

These qualitative insights illustrate the complex mechanisms through which conflicts worsen food security. The following section details the data and methodologies used to quantify the impact of conflicts on overall food security in African countries.

\section{Methods}
\subsection{Data}
This study has selected two main data sources which, appropriately combined, allow to estimate the direct impact of armed conflicts on FS. These are the FS assessments data regularly published by the Famine Early Warning Systems Network (FEWSNET), and conflict data collected by The Armed Conflict Location Event Data Project (ACLED). \\
The subsequent subsections provide a detailed explanation of each data source, along with a comprehensive account of the methodology employed to integrate them for our analysis.

\subsubsection{FEWSNET data}
Famine Early Warning Systems Network (FEWSNET) is an entity established by the United States Agency for International Development (USAID) in 1985 with the specific goal to fight hunger, and is today considered the gold standard of food security assessments and projections.
Its primary contribution to addressing hunger lies in the regular publication of Food Security Outlook reports. Since 2016, these reports have been issued three times per year, specifically in February, June, and October. Reports are produced for each country where FEWSNET operates (Fig 1), and provide comprehensive insights into the current food security situation, its underlying causes, and future projections. Additionally, the reports outline assumptions about how these conditions will evolve and their anticipated impact on food security in the coming six months. 

\newpage
\begin{figure}[ht]
\centering
\includegraphics[width=0.8\textwidth]{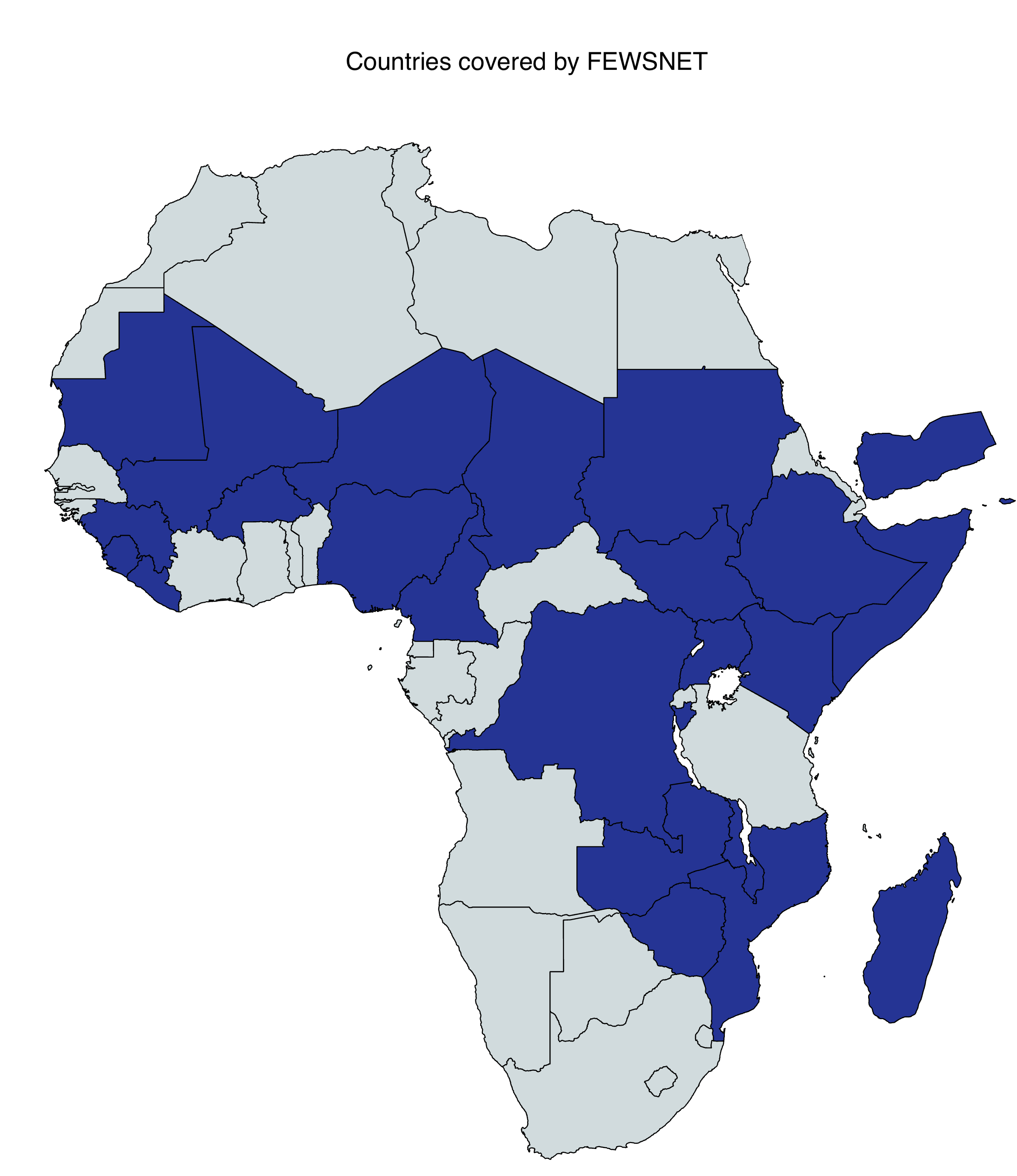}
\captionsetup{labelfont=bf}
\caption{Countries in blue are those for which FEWSNET generates regular assessments, hence included in this study.}
\end{figure}

FEWSNET employs a comprehensive methodology in the creation of its reports, a key aspect of which involves the assessment of FS in various geographical areas being surveyed. This assessment utilises the Integrated Food Security Phase Classification (IPC) scale, a five-class system designed to provide a detailed understanding of food security levels. The IPC scale spans from Class 1, signifying ``minimal'' food insecurity, to Class 5, denoting the most severe condition, ``Famine''. Throughout this analysis, the terms ``phase'' and ``class'' will be used interchangeably to refer to the different levels on the IPC scale. The table below  (Fig 2) provides a detailed description of each phase.

\newpage
\begin{figure}[ht]
\centering
\includegraphics[width=\textwidth]{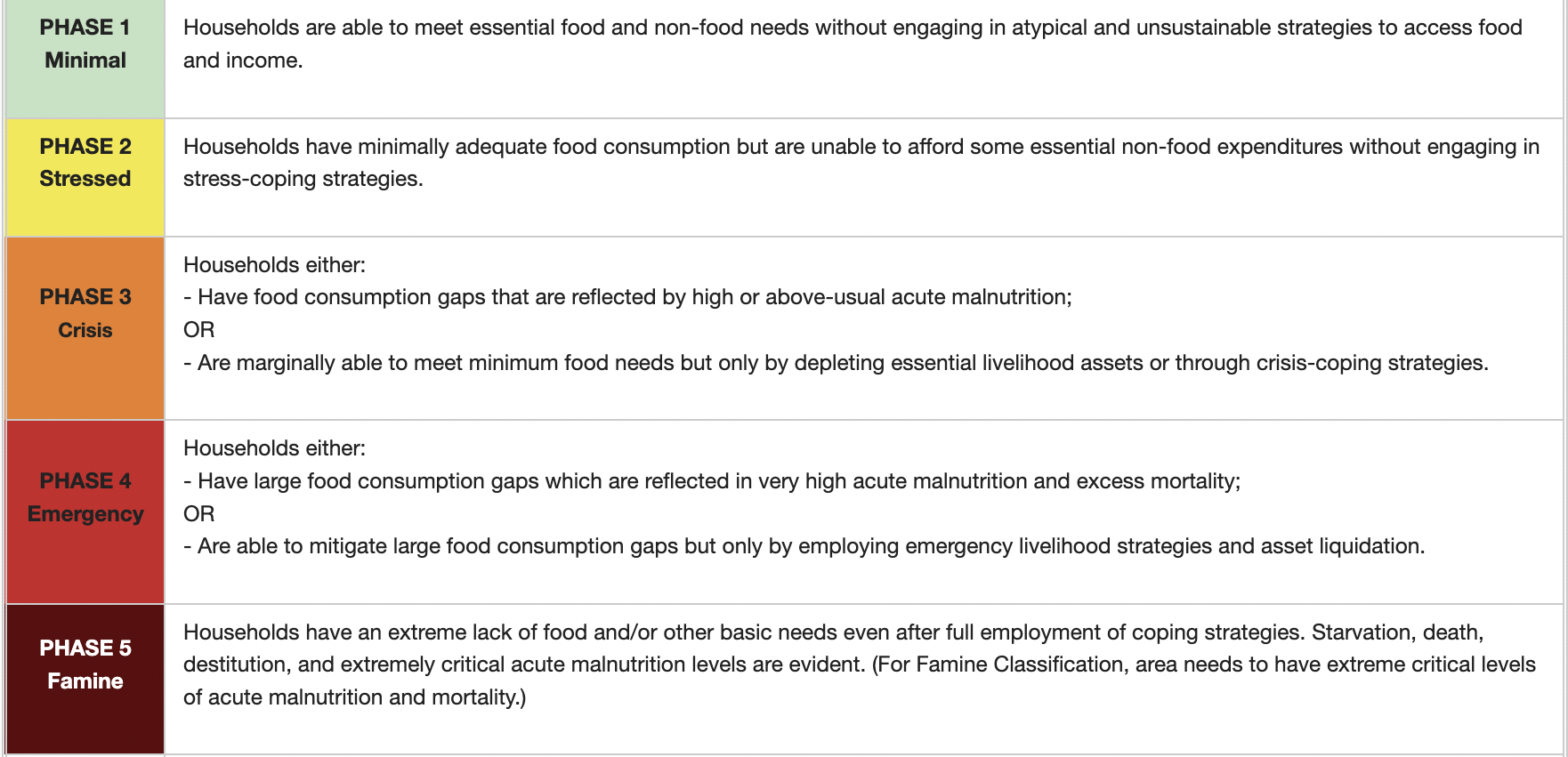}
\caption{Source: fews.net}
\end{figure}

With each publication cycle, FEWSNET releases the raw data behind each report, enabling researchers to conduct independent work on the FS topic. The key datasets made available, and those utilised in this study are the following:

\begin{itemize}
  \item \textbf{Food Security Classification}: Provides the IPC phase classification for each geographical area. Specifically, for this study we used the Current Situation (CS) assessment for the state of FS at the time of the reports' publication. Each file consists of one row for each of the 5 IPC FS classes, and each class has a set of polygons associated to them, which allows to draw the different areas by their given state of FS class;
  
  \item \textbf{Administrative Boundaries}: shapefiles containing polygons with information about sub-national level administrative boundaries; 

\end{itemize} 

This study uses data from reports published since 2016, when FEWSNET transitioned from a quarterly publication schedule to the current practice of releasing reports three times per year. By utilising data since 2016 we ensures consistency in the interval between reports over time.

\subsubsection{ACLED data}
The Armed Conflict Location Event Data Project (ACLED)\cite{ACLED} provides possibly the best curated source of information about conflicts; from terrorist attacks to traditional wars, to political unrest and peaceful protests, ACLED provides easy access to information such as location, number of fatalities, actors, dates, as well as a comprehensive free text description of each event. This data in independently collected and updated daily, with complete geographical coverage, and a clear explanation of the violent event that occurred. 

\newpage
\subsection{Methodology}
\begin{figure}[ht]
\centering
\includegraphics[width=1\textwidth]{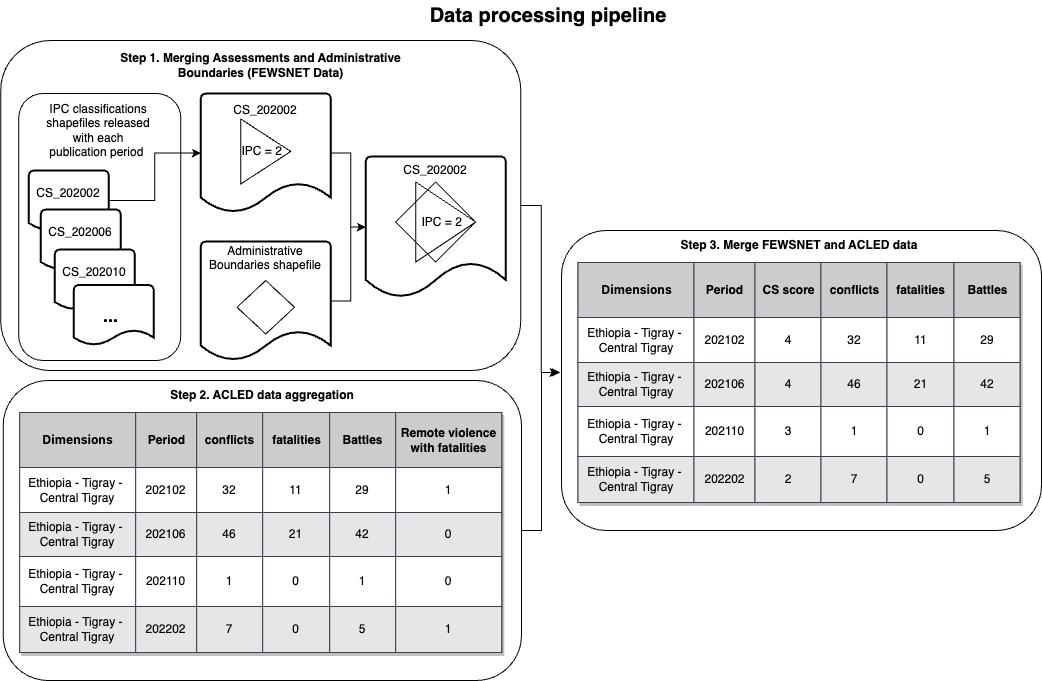}
\caption{Diagram of the data processing pipeline: in Step 1 we merge Administrative Boundaries with each published FS assessment; Step 2 aggregates ACLED data, counting conflicts by Administrative boundaries and event period; Step 3 combines all the data into a single table, covering assessments and count of conflicts over time and geographies}
\end{figure}

To precisely evaluate the impact of conflicts on the state of Food Security (FS), it is imperative to construct a robust framework capable of facilitating a comprehensive correlation analysis, as well as the training, evaluation, and comparison of Machine Learning (ML) models. Establishing this framework, we amalgamated the data outlined in the preceding section to create a geospatially consistent dataset. This dataset serves as the foundation for the analysis presented in this study, across both temporal and geographical dimensions. Ensuring spatial consistency involved mapping spatial data from each Famine Early Warning Systems Network (FEWSNET) CS report onto the latest administrative boundaries, as provided by FEWSNET. These boundaries, known for their inherent stability, only change over extended periods, ensuring a reliable baseline for comparison. \\

We later aggregated data from The Armed Conflict Location Event Data Project (ACLED), counting the number of conflicts (both overall and by type) and the number of fatalities. This ACLED data was then merged with FEWSNET data based on period and administrative boundaries. To observe the impact of conflicts occurring in the previous three months on the subsequent state of FS, we introduced a three-month lag between ACLED and FEWSNET data. This deliberate lag allows for a nuanced analysis of the temporal dynamics between conflict events and FS outcomes. The resulting data was then used to perform both the correlation analysis as well as the ML modelling work.\\

The following subsections provide a detailed explanation of the computational steps we have taken to process the data, including data merging and consistency assessments of FS. We outline the methodology used, the rationale behind each step, and how these steps contribute to ensuring data reliability and accuracy. Additionally, the logic used to create the rule-based models is explained.

\subsubsection{Data processing}
The key step, was to combine the data from the administrative boundaries and the Food Security Classification files. This was accomplished by intersecting each of the polygons containing the IPC classification for each period of publication with the polygons containing administrative boundaries, creating more, smaller areas containing information from both sources (Fig 3, Step 1)\footnote{Total number of merges is 189. This is equal to 7 years of data * 3 publications per year * 3 files per publication (CS, ML1, ML2) * 3 geographical areas (SA, EA, WA) each merged administrative boundaries.}.

Since administrative boundaries seldom change, they provide a stable and granular geo-spacial dimension for comparisons over time; We have used the most recent files (as of September 2023) as fixed point for the mapping. \\

The figures below show and example of the raw data contained in each dataset. 

\newpage
\begin{figure}[ht]
  \centering
  \subfloat{\includegraphics[width=0.45\textwidth]{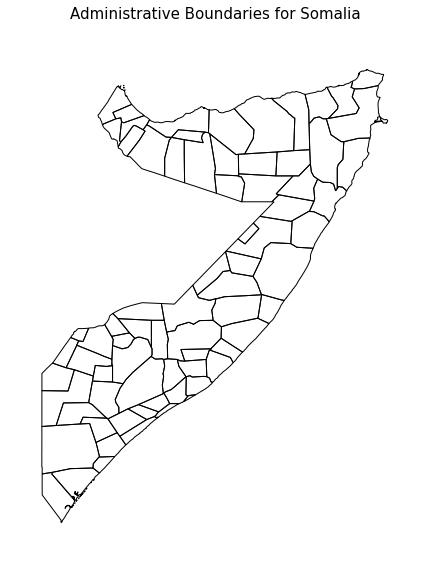}}
  \subfloat{\includegraphics[width=0.47\textwidth]{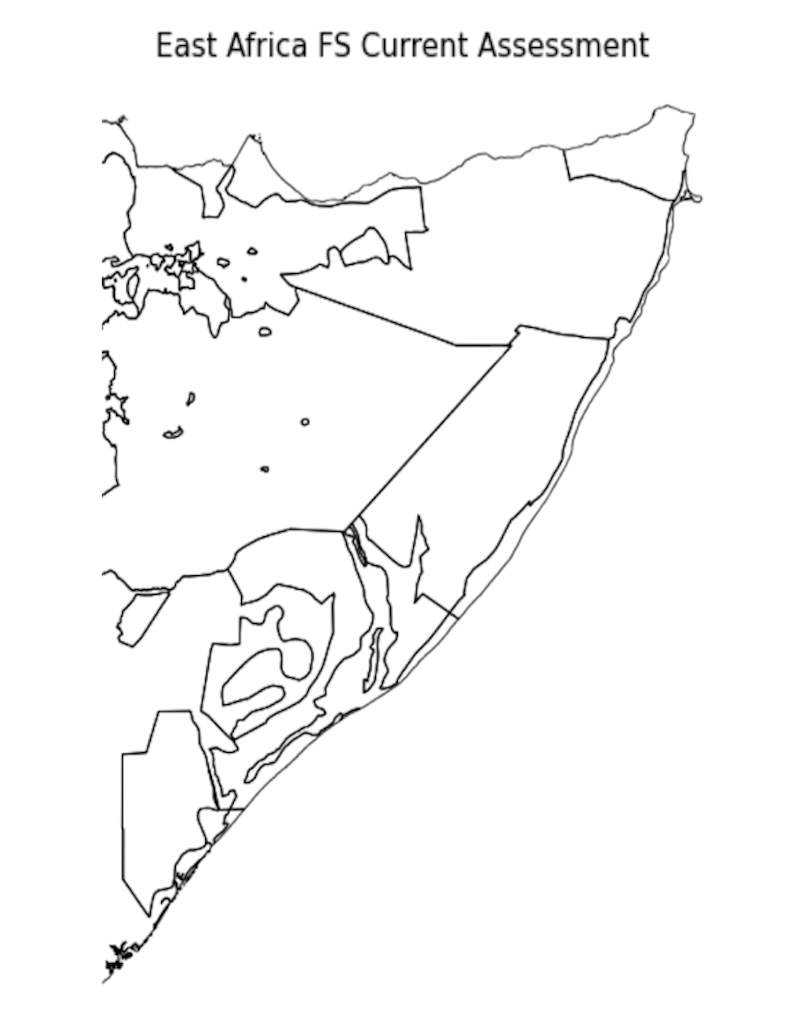}}
  \hfill
  \caption{Example of visualisation of Administrative Boundaries and FS assessment in Somalia from FEWSNET shape files.}
\end{figure}

After computing the intersection, we have removed polygons with an area smaller than 0.005; these very small polygons can occur due to the non-perfect overlapping between the original shapes, and can cause duplication in the FS-Admin combinations; removing such small areas resolved this problem without eliminating any significant amount of land coverage\footnote{This removes less than 0.01\% of the total area.}.\\

The output of this process are new shapefiles, covering the same geographical area as the original Food Security Classification files from FEWS, but with an increased number of polygons, each representing the overlap between FS classification and administrative boundaries (Fig 5).\\

\begin{figure}[ht]
\centering
\includegraphics[width=0.5\textwidth]{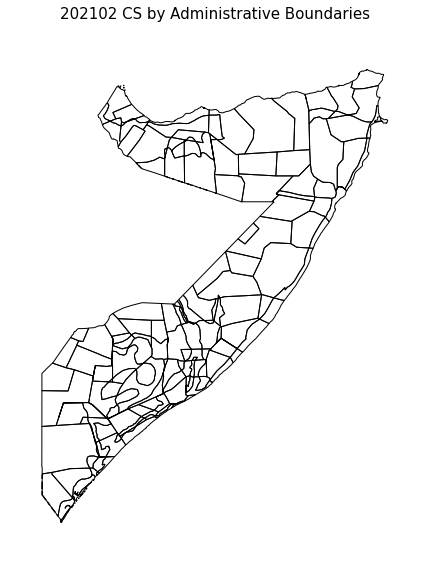}
\caption{Visualisation of the final output dataset, result of combining Administrative Boundaries and FS assessment shape files published by FEWSNET (CS for February 2021)}
\end{figure}

During this step we also removed duplicates from the data, aggregating the data by each FS-Admin combination and, in case any duplicate occurred, we took the highest (worse) FS score for each of these areas. The resulting dataset is as granular as district level (ADMIN2), which is the same level of spatial granularity at which ACLED collects data\footnote{This operation affects between 3 and 5\% of the dataset (depending on the period), where the same Admin-Livelihood have different FS scores.}. Next, we processed ACLED data (Fig 3, Step 2). We used the event date to extract the ``period'' in the same format used by FEWSNET (year-month, i.e. 202202). We then computed the count of conflicts and fatalities at the administrative boundaries and period level, both overall and by conflict type. \\

Lastly, we merged FEWSNET and ACLED data by joining the two datasets on country, region, district names, and periods (Fig 3, Step 3). In doing this, we introduced a three-month lag between ACLED and FEWSNET data. For example, for the FEWSNET publication period of 202210, we counted the number of conflicts occurring between 202207 and 202209. \\

This method of aggregating data at the district level allowes for a spatially granular analysis of the relationship between the state of FS and different types of violent events, and offers several advantages over raster-based approaches commonly used in the literature (e.g., Backer and Billing, 2021). Firstly, it aligns with FEWSNET's prediction granularity. FEWSNET's FS assessment and prediction involves using a combination of socioeconomic data, weather forecasts, and surveys, which are made at the district level. This approach enables a straightforward usage of FEWSNET data. 

Secondly, this data aggregation method simplifies data interpretation and communication. By grouping information to the administrative boundaries level, it provides a more intuitive way to convey results and findings to policymakers and stakeholders. 

Lastly, this level of aggregation will facilitate the following steps of this study, which involves generating FS prediction using conflicts data. It would be extremely challenging, if at all possible, to map the data used to make predictions to specific areas withing a raster grid; the current approach allows to map events to specific districts, streamlining and simplifying the methods for predictions. 

\subsubsection{Benchmark models}
For the ML part of this work, we have established four distinct baselines. Three of them are created by applying simple heuristics on top of FEWSNET's data, and are designed to provide insights on potential seasonal and recurring states of FS. The fourth model actually uses ML to make predictions, using historical states of FS as variables. \\

Below is a detailed description of each baseline we have computed:
\begin{itemize}

\item \textbf{Previous Period's IPC Score (PPS)}: This first model assumes a scenario of no change in food security conditions and simply replicates the Integrated Food Security Phase Classification (IPC) score from the preceding period. It provides insight into the stability of food security.

\item \textbf{Same Period Last Year (SPLY)}: Recognizing the influence of seasonal factors on food security, this model considers the IPC score from the same period in the previous year. By accounting for seasonality, it offers a valuable perspective on the cyclical nature of food security predictions.

\item \textbf{Maximum of Previous Two Periods (Max-2PP)}: This model selects the highest (worst) IPC score from the previous two periods. This approach takes into account recent trends and variations in food security, highlighting potential deteriorations.

\item \textbf{Classifier using historical CS scores (CHS)}: This ML model was trained using exclusively historical FS data included in FEWSNET reports, such as the geographical region, the number of transitions between crisis and non-crisis, and lagged CS scores.

\end{itemize} 

The following section will provide a detailed description of the different metrics that we used in the correlation analysis and in the comparison of the different prediction methods.

\section{Metrics}
In this section, we discuss the metrics used to evaluate both the food security conditions and the machine learning models. These metrics provide insights into model performance and the stability and trends in food security, allowing for a comprehensive analysis of our results.

\subsection{Correlation analysis key metric}
Correlation analysis plays a critical role in understanding the relationships between variables, which, in this study, involves exploring how conflict dynamics influence FS conditions in Africa. To achieve this, Spearman correlation is utilized as a key metric to measure the strength of relationships between conflict occurrences and FS indicators. 

\begin{itemize}

\item \textbf{Spearman correlation} is a non-parametric measure assessing the strength and direction of the monotonic relationship between two variables. Ranging from -1 to 1, positive values indicate a monotonically increasing relationship, negative values a decreasing one, and 0 signifies no monotonic correlation. Spearman correlation is robust to outliers and is suitable for ordinal or non-normally distributed data, providing insights into relationships between conflict occurrences and food security (FS) conditions.

\end{itemize} 

\subsection{Machine learning model metrics}

Part of the the objective of this study is assess how integrating conflict data can improve the predictive performance of models estimating the state of FS. In order to fairly compare the performance of different methods, five main metrics will be used: accuracy, precision, recall, F1 and AUC. These are all metrics commonly used to measure the performance of ML classification problems, and they are defined as follows:

\begin{itemize}

\item \textbf{Accuracy} is a measure of how well the classification model correctly predicts the target variable. It is calculated by dividing the number of correct predictions by the total number of predictions. In this study, accuracy is calculated as the proportion of geographical areas where the models correctly predicted the Integrated Phase Classification (IPC) scale, ranging from 1 (minimal food insecurity) to 5 (famine). While accuracy provides a general overview of the model's correctness, it can be misleading in imbalanced datasets. This is particularly relevant in FS analysis where severe crisis situations (IPC 4-5) are thankfully the minority of total cases, meaning that high accuracy might still overlook many areas at risk if the model predominantly predicts non-crisis situations.

\item \textbf{Precision} focuses on the model's positive predictions, specifically its ability to avoid false alarms regarding food security crises. Precision is calculated as the ratio of true positive predictions (correctly identified areas in crisis) to the total predicted positives (areas identified as in crisis by the model). High precision is crucial for effectively deploying limited resources. In the context of food security, false positives could result in unnecessary allocation of aid, which is especially costly given the limited resources available for intervention across affected regions.

\item \textbf{Recall}, also known as sensitivity or true positive rate, measures the model’s ability to correctly identify areas truly in crisis. It is calculated as the ratio of true positives to the actual positives (areas that are genuinely facing a crisis). Recall is particularly valuable in this study because missing an area experiencing a crisis (a false negative) can be critical, potentially leading to insufficient intervention. Thus, maximizing recall ensures that the model captures as many crisis situations as possible, which is vital for timely and targeted food aid.

\item \textbf{F1 score} is the harmonic mean of precision and recall, offering a balanced measure when both metrics are important. In the context of this research, the F1 score helps evaluate the model's ability to predict food security levels effectively, ensuring that it balances the need to avoid false positives with the need to minimize false negatives. A high F1 score indicates that the model is effectively distinguishing between areas in need of immediate support and those not at risk, thus enhancing the efficient allocation of resources.

\end{itemize} 
In the next section, we describe the results we have obtained from both the correlation analysing and the machine learning modelling of conflicts to predict FS.

\section{Results}
As previously mentioned, our analysis was conducted in two different steps. In the first step, we performed a correlation analysis between the lagged count of conflicts and fatalities; in the second step we compared an ML model trained using features extracted from conflicts data against simple rule-based models, as well as a model exclusively trained on historical FS data. The following two sections will present the results of each step.

\subsection{Correlation analysis}
We performed our correlation analysis at three different levels of spatial granularity: country, region, and district. This multi-level approach allowed us to explore the relationships between conflict events and food security outcomes at varying geographic scales, providing a comprehensive understanding of the dynamics involved. At the country level, correlations provide an overview of broad trends, highlighting national patterns between conflict and FS. This is particularly valuable for policymakers as it offers a high-level perspective for strategic planning and allocation of resources. Analysing correlations at the regional and district levels provides more nuanced insights. These finer geographic distinctions help reveal localized areas of vulnerability that may be obscured at broader scales. Such insights are crucial for tailoring interventions effectively, as they highlight specific regions where FS is more sensitive to conflict dynamics. In each iteration, after computing the correlations, we only visualized relationships that were statistically significant\footnote{Defined as p-value < 0.05}. This method ensures that we focus on meaningful relationships that can guide both policy and operational decisions, thereby adding practical value to our findings.\\

At the country level (Fig. 6), significant correlations higher than 0.4 emerge in six nations: Burkina Faso, Mali, Kenya,  Ethiopia, Sudan and Malawi. We qualitatively verified the results by cross-referencing countries with high correlations between food security and conflicts against news articles and official UN reports, specifically checking whether conflicts were mentioned as key causes of food crises. In each case, we found alignment between our model's findings and the documented causes. In the country showing the strongest correlation, Burkina Faso (BF), a discernible upward trend in violent conflicts has been observed since 2018, leading to significant displacement and disrupting food and water supplies. Towns such as Djibo in the Sahel region have endured prolonged sieges, resulting in severe food insecurity\cite{IDMC_NRC_BurkinaFaso}. Mali has been afflicted by violence caused by ethnic tensions, regional disputes, and the presence of terrorist groups, contributing to political instability since 2012\cite{HRW2022Mali}. The situation has further worsened since 2020 following several coup d’état\cite{CrisisGroupMali2022}. In Kenya, where internal conflicts have persisted since the transition to multi-party politics in the 1990s, there has been a recent escalation in atrocities. This surge is fueled by the commercialization and politicization of cattle rustling, exacerbating ethno-political competition\cite{PeaceCameHerdersKenya}. 

\begin{figure}[ht]
\centering
\includegraphics[width=\textwidth]{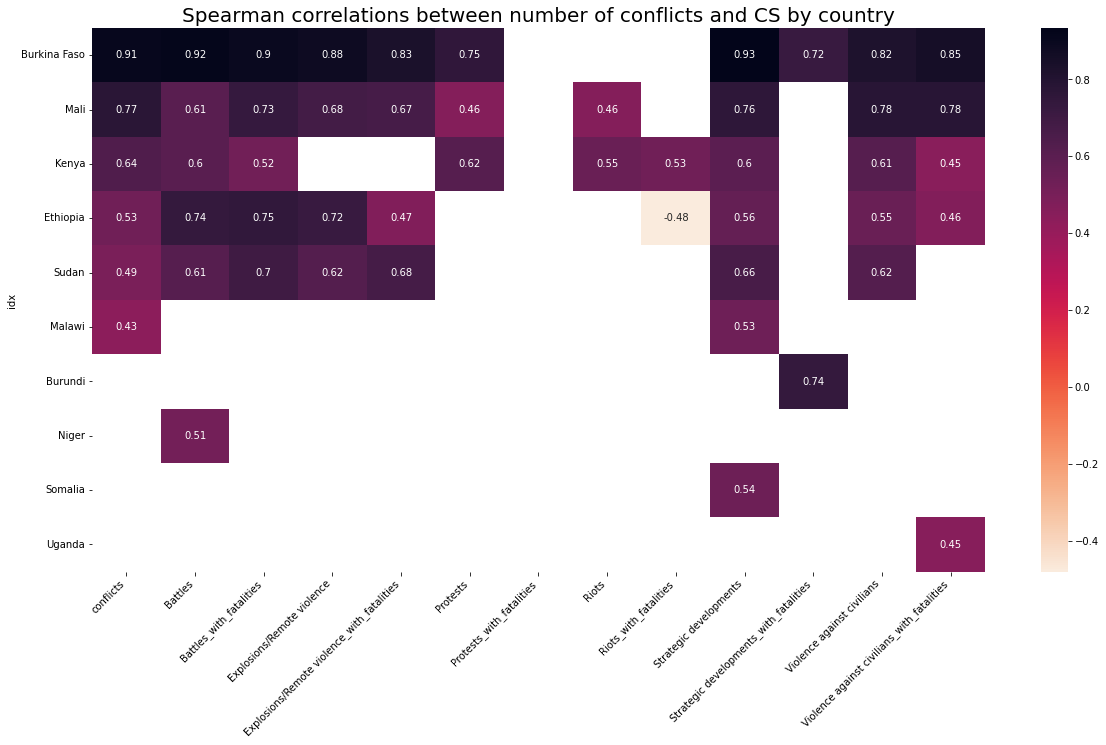}
\caption{Spearman correlation between the lagged count of conflicts and CS at country level (February 2016 - February 2024). Only correlations that are statistically significant (p-value \textless{} 0.05) are shown.}
\end{figure}

At the regional level (Fig 7), for those countries that were already showing correlations at national level, we can clearly see the localised causes of the link between FS and conflicts. Notably, in Burkina Faso (BF), the Sahel and surrounding regions emerge as the primary contributors to the observed correlation. In Kenya, the Turkana and Marsabit regions are the main drivers of how FS is impacted by violence. For some countries, the lower granularity reveals dynamics that where hidden by aggregating data at a country level. Amongst the highest correlated regions, we find several ones from Nigeria, Sudan, Mozambique, and Niger, where country level correlations where not significant. For instance, in Nigeria's Katsina, Kaduna, and Niger state, a recent surge in kidnappings has led to the displacement of local farmers. Those who remain are grappling with a scarcity of labor for agricultural activities, stemming from the pervasive fear of abduction\cite{ForbesNigeriaBanditry2022}. \\

\newpage

\begin{figure}[ht]
\centering
\includegraphics[width=\textwidth]{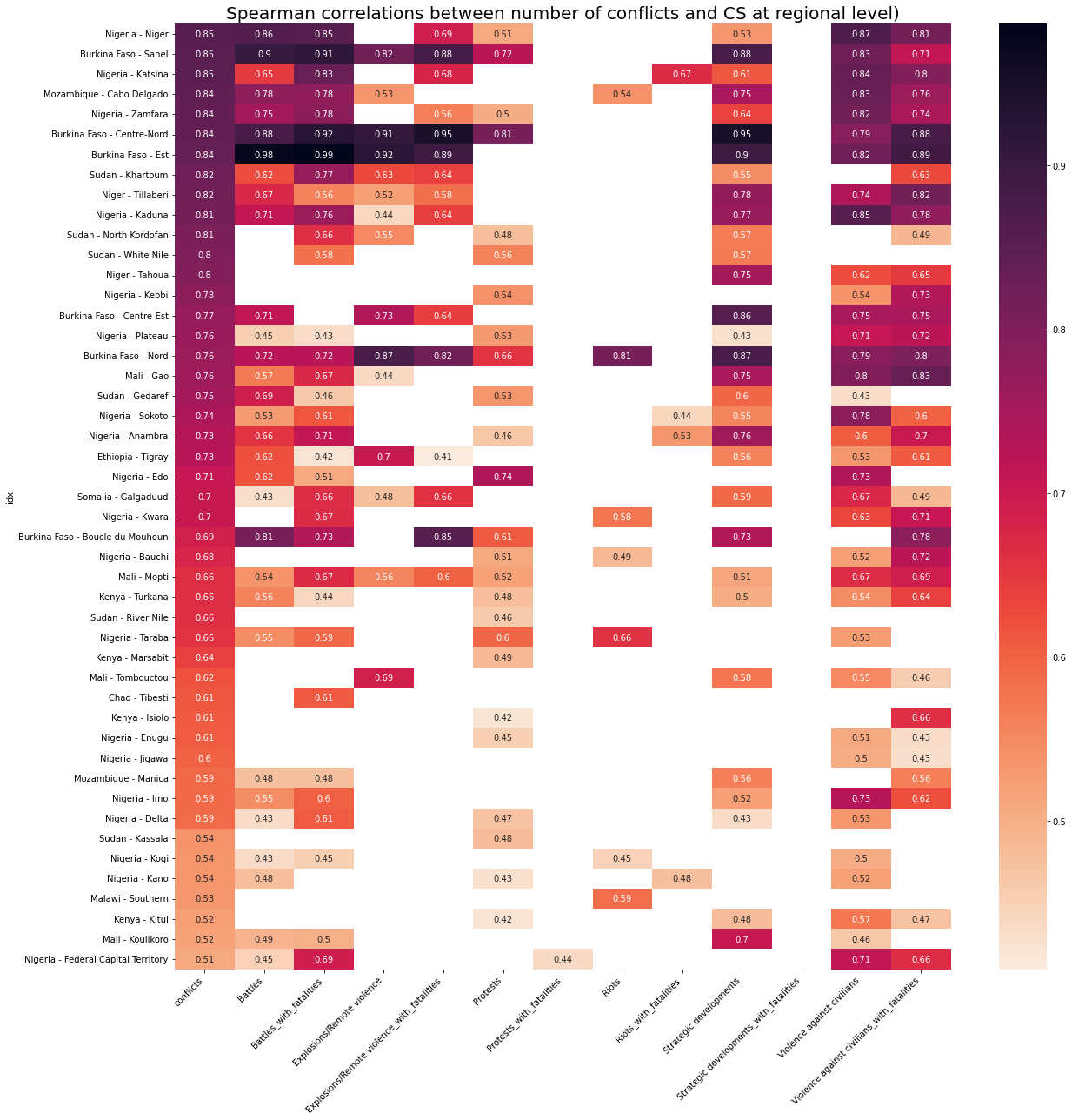}
\caption{Spearman correlation between the lagged count of conflicts and CS at region level (February 2016 - February 2024). Only correlations that are statistically significant (p-value \textless{} 0.05) are shown.}
\end{figure}

In exploring the district level (Fig. 8), no new regions emerge that did not already exhibit significance at higher levels of aggregation. Nevertheless, this finer granularity does offer valuable insights into localized conflicts. For instance, it allows us to pinpoint specific districts and cities (ADMIN2) that are most impacted by conflict. This detailed information can prove instrumental for enhancing the accuracy of food security (FS) predictions, especially when employing machine learning (ML) models.\\

\newpage
\begin{figure}[ht]
\centering
\includegraphics[width=\textwidth]{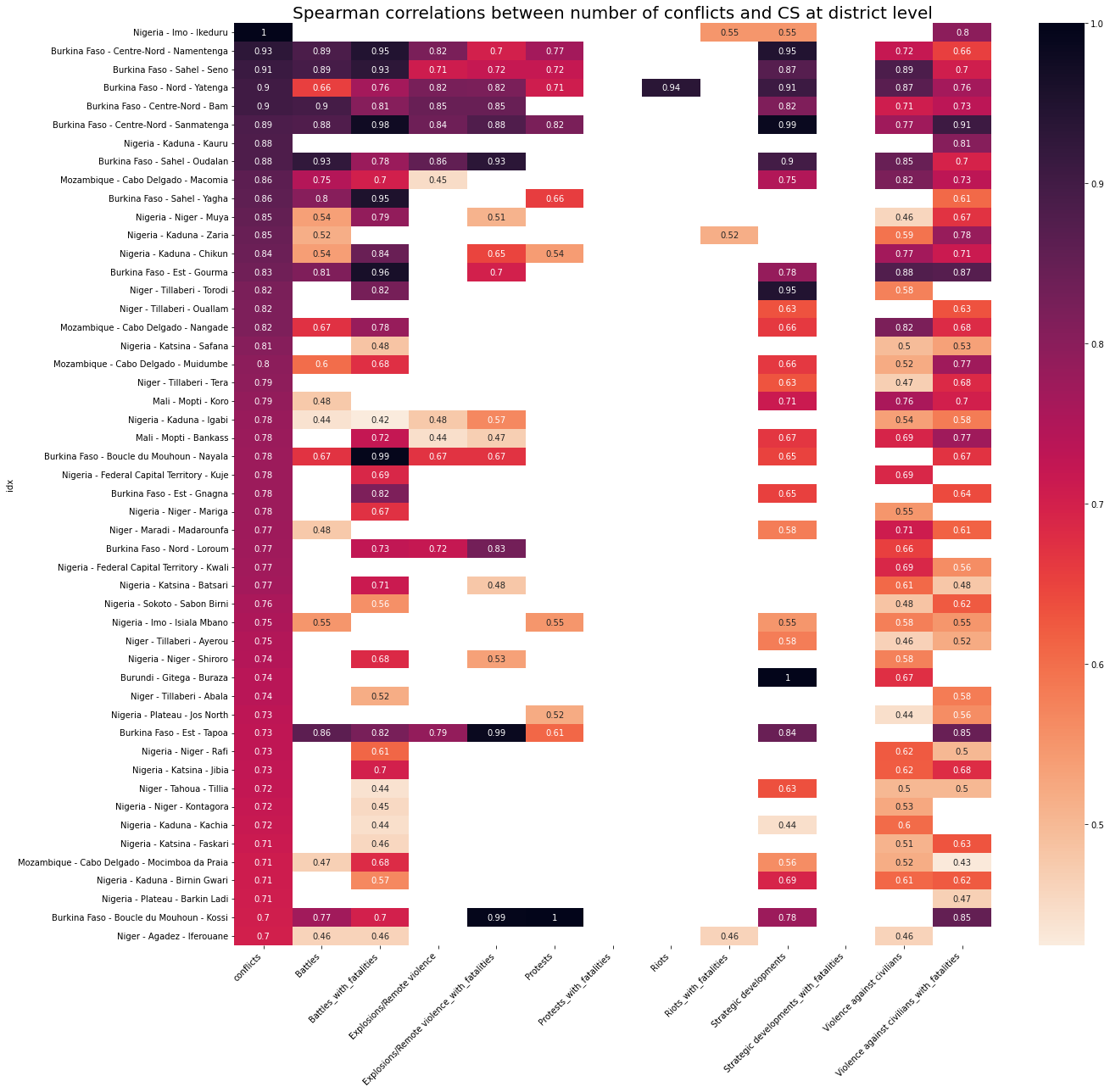}
\caption{Spearman correlation between the lagged count of conflicts and CS at district level (February 2016 - February 2024). Only correlations that are statistically significant (p-value \textless{} 0.05) are shown.}
\end{figure}

In the following subsection we will show the effect of using features extracted from conflict data to train ML models to predict FS.

\subsection{Modelling conflicts to predict FS} 
Our study's second phase centered on evaluating the impact of incorporating conflict data into FS predictions. To ensure a comprehensive assessment, we employed the CHS classifier, augmenting the dataset with conflict-related features. These supplementary features comprised the three-months lagged counts of conflicts and fatalities, along with the cumulative sums of these metrics over the preceding two years.\\

The earlier correlation analysis provided insights into where food security and conflict were statistically linked, highlighting broad areas where these two factors were connected. However, this phase moved beyond correlation by demonstrating how the use of conflict data directly improved the predictive performance of our ML models. Specifically, adding conflict features resulted in an overall improvement in model accuracy, precision, and recall, effectively showing that the inclusion of conflict data increased the model's ability to correctly identify areas at risk of food insecurity. This improvement highlights the value of incorporating conflict information not just for understanding relationships, but for enhancing the accuracy and reliability of food security forecasts, ultimately enabling more proactive interventions. \\

To achieve this, we employed Logistic Regression (LR) and Random Forest (RF) due to their specific strengths. LR was selected for its simplicity and interpretability, allowing us to easily understand the relationship between predictors and outcomes, which is crucial for deriving actionable insights in the context of food security. RF algorithm was chosen for its ability to effectively capture non-linear relationships in the data, its ease of interpretability through the display of feature importance (as shown in Fig. 9), and its robustness against overfitting. This was particularly important given the complexity of interactions between conflict and food security.

We did not attempt to use deep learning models due to their nature as black-box techniques, which would not provide insights into the specific contribution of conflict features to the model's predictions. Moreover, our dataset did not have a large number of features that would justify using such complex models. \\

The summarized performance of each model is presented in the table below (Table 1).

\clearpage
\begin{sidewaystable}[ht]
    \makebox[\textwidth]{%
        \begin{minipage}{0.7\textwidth}
        \centering
        \large \textbf{Performance Comparison of Different Machine Learning Models for Historical and Conflict Data}
        \end{minipage}
    }
    \vspace{1cm} 

    \renewcommand{\arraystretch}{1.3}

   \begin{tabular}{>{\centering\arraybackslash}m{2cm} >{\centering\arraybackslash}m{1cm} p{8cm} cccc}
    \toprule
    \textbf{Type} & \textbf{Index} & \textbf{Model} & \textbf{Test Accuracy} & \textbf{Test Precision} & \textbf{Test Recall} & \textbf{F1} \\
    \midrule
    Rule-based & 1 & PPS & 0.734 & 0.753 & 0.734 & 0.735 \\ \hline
    Rule-based & 2 & SPLY & 0.705 & 0.719 & 0.705 & 0.710 \\ \hline
    Rule-based & 3 & Max-2PP & 0.788 & 0.810 & 0.788 & 0.789 \\ \hline
    ML-based & 4 & LogisticRegression(class\_weight='balanced'), CHS & 0.742 & 0.754 & 0.742 & 0.744 \\ \hline
    ML-based & 5 & RandomForestClassifier(class\_weight='balanced', n\_estimators=200, n\_jobs=-1, random\_state=5), CHS & 0.748 & 0.762 & 0.748 & 0.752 \\ \hline
    ML-based & 6 & LogisticRegression(class\_weight='balanced'), including conflict features & 0.757 & 0.771 & 0.757 & 0.760 \\ \hline
    ML-based & 7 & RandomForestClassifier(class\_weight='balanced', n\_estimators=200, n\_jobs=-1, random\_state=5), including conflict features & 0.763 & 0.779 & 0.763 & 0.767 \\ \hline
    \bottomrule
    \end{tabular}

    \caption{Summary table showing the performance of different models, three rule-based (PPS, SPLY, Max-2PP) and two ML-based (Logistic Regressions with and without integrating conflict-related data).}
    \label{tab:performance}
\end{sidewaystable}

\clearpage

Both ML models outperformed the heuristic-based models across all evaluation metrics. Notably, the model incorporating conflicts-related features exhibited a 1.5\% higher accuracy compared to its counterpart trained solely on historical FS-related features.

Furthermore, the graph below (Fig. 9) illustrates that while traditional FS data such as the last period's score and the number of transitions in FS for a given region yielded robust predictive signals, the four conflict-related features integrated into the model ranked among the top ten most influential variables. This underscores the advantages of incorporating such data into FS prediction models.

\begin{figure}[ht]
\centering
\includegraphics[width=1\textwidth]{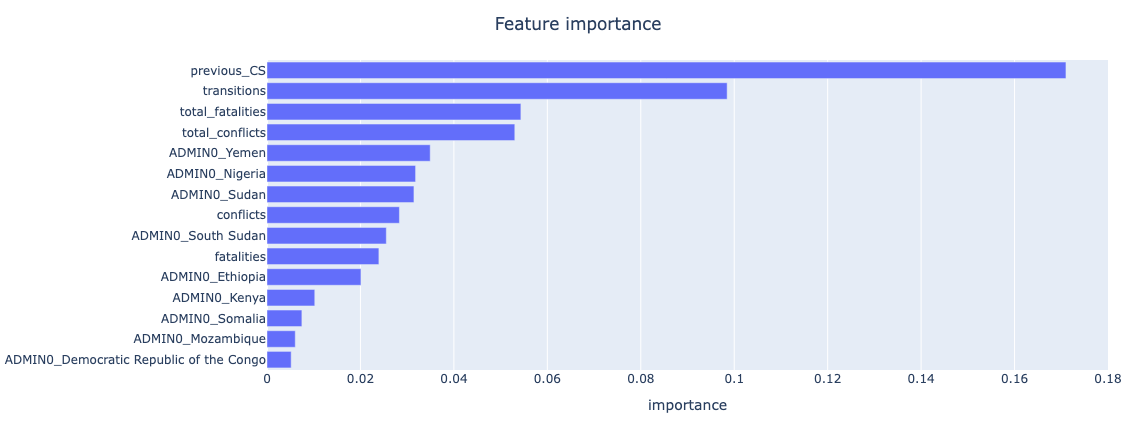} 
\caption{Feature importance of the Logistic Regression model that includes conflict-related data. Features created from conflict data rank high on importance.}
\end{figure}

\section{Discussion}
Our analysis has demonstrated that incorporating conflict data significantly enhances the accuracy of models predicting FS outcomes. The correlation analysis at different spatial levels revealed that conflict events are highly associated with food insecurity in specific countries and regions, such as Burkina Faso, Kenya, and parts of Nigeria. These findings are consistent with previous research (Kemmerling et al. 2022; Brück et al. 2016) that highlights the disruptive effects of conflict on local food systems, markets, and community resilience. However, to our knowledge, this study is the first to quantify the impact of conflict on food security in a rigorous, quantitative manner across multiple geographic levels. This approach allows us to identify specific regions and districts most affected by conflict, offering valuable insights for targeted interventions. \\

The improved performance of machine learning models with conflict-related features underscores the value of integrating such data into predictive frameworks. The inclusion of these features not only boosted model accuracy but also emphasized the importance of conflict dynamics as critical drivers of FS. This finding aligns with the hypothesis that conflicts disrupt agricultural production, market access, and food distribution networks, thereby contributing to worsening the state of FS. \\

Despite these promising results, there are several limitations to consider. The use of lagged counts of conflict events assumes a linear temporal relationship between conflicts and food security outcomes, which may not always capture the complexity of these interactions. Moreover, the data used for this study may be subject to reporting biases or inaccuracies, particularly in conflict zones where data collection is challenging. \\

Overall, our findings suggest that incorporating conflict data into FS prediction models offers a promising approach to improving early warning systems and informing policy decisions. By providing the first quantitative assessment of the link between conflict and FS, this study contributes significantly to the broader effort to develop more responsive and effective interventions in conflict-affected areas.

\section{Conclusions}

This paper presents the first quantitative assessment of the impact of incorporating conflict data into predictive models for food security (FS). Our correlation analysis reveals that conflicts are significant drivers of food insecurity, particularly in specific countries and regions across Africa. By employing a multi-level spatial analysis, we were able to identify both broad and localized relationships between conflicts and FS outcomes, highlighting areas where conflict plays a pivotal role in exacerbating food insecurity. \\

Our modeling exercise demonstrates a clear enhancement in predictive accuracy when conflict-related features are included in machine learning models. Specifically, models that integrated conflict data achieved a 1.5\% higher accuracy compared to models trained solely on historical FS features. This finding is crucial as it quantitatively shows how incorporating conflict information enhances the capacity of predictive models to accurately identify areas at risk of food crises, thereby providing a more reliable basis for proactive interventions.

The key contributions of this study include:
\begin{itemize}
    \item \textbf{Quantitative Insights into Conflict Impact}: We provide a robust quantitative evaluation of the impact of conflicts on food security, using correlation analysis and predictive modeling to show the value of incorporating conflict data.
    \item \textbf{Predictive Improvement through Data Integration}: By integrating conflict-related features, our ML models outperformed rule-based approaches, demonstrating the importance of data-driven methods in enhancing food security predictions.
    \item \textbf{Actionable Insights for Policymaking}: The findings underscore the role of conflicts as critical factors influencing FS, providing evidence to inform policy decisions and enable more targeted interventions in conflict-affected regions.
\end{itemize}

Despite these promising outcomes, our approach also highlights areas for future research. The reliance on lagged counts of conflict events assumes a linear temporal relationship between conflicts and FS outcomes, which may not fully capture the complex and potentially nonlinear dynamics at play. Future research could address these limitations by exploring more advanced modeling techniques, such as temporal or spatiotemporal models, that can better capture the intricacies of conflict impact over time. \\

Additionally, our study used Logistic Regression and Random Forest models for their interpretability and effectiveness. However, with access to larger datasets, future work could explore the use of more sophisticated models, such as deep learning techniques, which may provide further gains in predictive performance. Incorporating other drivers of food insecurity, such as climate data or socio-economic indicators, could further enhance the robustness of FS predictions and allow for more comprehensive early warning systems. \\

Overall, this study contributes significantly to the field of food security analysis by demonstrating the tangible benefits of integrating conflict data into predictive models. These insights are critical for developing more effective, evidence-based policies and interventions that address food insecurity in conflict-affected regions, ultimately helping to alleviate hunger and build resilience in vulnerable communities.

\clearpage 

\bibliography{sources}

\begin{thebibliography}{10}
\expandafter\ifx\csname url\endcsname\relax
  \def\url#1{\texttt{#1}}\fi
\expandafter\ifx\csname urlprefix\endcsname\relax\def\urlprefix{URL }\fi
\expandafter\ifx\csname href\endcsname\relax
  \def\href#1#2{#2} \def\path#1{#1}\fi

\bibitem{WFP-GlobalReport2024}
{World Food Programme (WFP)}, \href{https://www.wfp.org/news/global-report-food-crises-acute-hunger-remains-persistently-high-59-countries-1-5-people}{{2024 Global Report on Food Crises: Acute hunger remains persistently high in 59 countries with 1-in-5 people assessed in need of critical urgent action}}, accessed on September 29, 2024 (2024).
\newline\urlprefix\url{https://www.wfp.org/news/global-report-food-crises-acute-hunger-remains-persistently-high-59-countries-1-5-people}

\bibitem{kemmerling2022}
B.~Kemmerling, C.~Schetter, L.~Wirkus, The logics of war and food (in)security, Global Food Security 33 (2022) 100634.
\newblock \href {https://doi.org/10.1016/j.gfs.2022.100634} {\path{doi:10.1016/j.gfs.2022.100634}}.

\bibitem{Brown2020}
M.~Brown, D.~Backer, T.~Billing, et~al., \href{https://doi.org/10.1007/s12571-020-01041-y}{Empirical studies of factors associated with child malnutrition: highlighting the evidence about climate and conflict shocks}, Food Sec. 12 (2020) 1241--1252.
\newblock \href {https://doi.org/10.1007/s12571-020-01041-y} {\path{doi:10.1007/s12571-020-01041-y}}.
\newline\urlprefix\url{https://doi.org/10.1007/s12571-020-01041-y}

\bibitem{brown2021}
M.~Brown, K.~Grace, T.~Billing, D.~Backer, Considering climate and conflict conditions together to improve interventions that prevent child acute malnutrition, The Lancet Planetary Health 5 (2021) e654--e658.
\newblock \href {https://doi.org/10.1016/S2542-5196(21)00197-2} {\path{doi:10.1016/S2542-5196(21)00197-2}}.

\bibitem{bruck2016}
T.~Brück, et~al., The relationship between food security and violent conflict, Report to FAO (December 2016).

\bibitem{Andree2020}
A.~et~al, Predicting food crises, Policy Research Working Paper, World Bank Group (2020).

\bibitem{wfp2020}
{World Food Programme (WFP)}, Hunger, conflict, and improving the prospects for peace, Tech. rep., WFP (October 2020).

\bibitem{fao2022}
{Food and Agriculture Organization of the United Nations (FAO)}, Impact of the ukraine-russia conflict on global food security and related matters, Under the mandate of the Food and Agriculture Organization of the United Nations (FAO) (April 8 2022).

\bibitem{csis2022}
{Center for Strategic and International Studies}, The russia-ukraine war and global food security: A seven-week assessment, and the way forward for policymakers, April 15, 2022 (2022).

\bibitem{fao2022_Prices}
{Food and Agriculture Organization (FAO)}, \href{https://www.fao.org/worldfoodsituation/foodpricesindex/en}{Fao world food situation - food prices index}, accessed on February 26, 2024 (2022).
\newline\urlprefix\url{https://www.fao.org/worldfoodsituation/foodpricesindex/en}

\bibitem{Raleigh2020}
C.~Raleigh, H.~Nsaibia, C.~Dowd, The sahel crisis since 2012, African Affairs 120 (08 2020).
\newblock \href {https://doi.org/10.1093/afraf/adaa022} {\path{doi:10.1093/afraf/adaa022}}.

\bibitem{eu2022}
{European Civil Protection and Humanitarian Aid Operations}, \href{https://civil-protection-humanitarian-aid.ec.europa.eu/where/africa/sahel_en#facts--figures}{Eu civil protection and humanitarian aid - sahel}, accessed on February 26, 2024 (2022).
\newline\urlprefix\url{https://civil-protection-humanitarian-aid.ec.europa.eu/where/africa/sahel_en#facts--figures}

\bibitem{economist2021}
{The Economist}, \href{https://www.economist.com/leaders/2021/01/23/ethiopias-government-appears-to-be-wielding-hunger-as-a-weapon}{Ethiopia's government appears to be wielding hunger as a weapon}, accessed on February 26, 2024 (January 23 2021).
\newline\urlprefix\url{https://www.economist.com/leaders/2021/01/23/ethiopias-government-appears-to-be-wielding-hunger-as-a-weapon}

\bibitem{ACLED}
{ACLED}, accessed on August 17, 2024 (2024).
\newblock \href{https://acleddata.com/}{[link]}.
\newline\urlprefix\url{https://acleddata.com/}

\bibitem{IDMC_NRC_BurkinaFaso}
{Internal Displacement Monitoring Centre (IDMC) and Norwegian Refugee Council (NRC)}, \href{https://api.internal-displacement.org/sites/default/files/IDMC_NRC_Conflict%2C_displacement_and_food_security_in_Burkina_Faso.pdf}{Conflict, displacement, and food security in burkina faso}, accessed on April 27, 2024 (2023).
\newline\urlprefix\url{https://api.internal-displacement.org/sites/default/files/IDMC_NRC_Conflict%2C_displacement_and_food_security_in_Burkina_Faso.pdf}

\bibitem{HRW2022Mali}
H.~R. Watch, \href{https://www.hrw.org/world-report/2022/country-chapters/mali}{World report 2022: Mali}, accessed on April 27, 2024 (2022).
\newline\urlprefix\url{https://www.hrw.org/world-report/2022/country-chapters/mali}

\bibitem{CrisisGroupMali2022}
I.~C. Group, \href{https://www.crisisgroup.org/africa/sahel/mali/nord-du-mali-une-confrontation-dont-personne-ne-sortira-vainqueur}{Nord du mali: Une confrontation dont personne ne sortira vainqueur}, accessed on April 27, 2024 (2022).
\newline\urlprefix\url{https://www.crisisgroup.org/africa/sahel/mali/nord-du-mali-une-confrontation-dont-personne-ne-sortira-vainqueur}

\bibitem{PeaceCameHerdersKenya}
{ReliefWeb}, \href{https://reliefweb.int/report/kenya/peace-came-herders-supporting-peace-amidst-drought-food-insecurity-and-conflict-lessons-marsabit-county-kenya}{Peace came to the herders: Supporting peace amidst drought, food insecurity, and conflict - lessons from marsabit county, kenya}, accessed on April 27, 2024 (2023).
\newline\urlprefix\url{https://reliefweb.int/report/kenya/peace-came-herders-supporting-peace-amidst-drought-food-insecurity-and-conflict-lessons-marsabit-county-kenya}

\bibitem{ForbesNigeriaBanditry2022}
D.~Ewing-Chow, \href{https://www.forbes.com/sites/daphneewingchow/2022/02/28/in-nigeria-banditry-has-become-a-major-threat-to-food-security/?sh=5068ee3861c3}{In nigeria, banditry has become a major threat to food security}, accessed on April 27, 2024 (2022).
\newline\urlprefix\url{https://www.forbes.com/sites/daphneewingchow/2022/02/28/in-nigeria-banditry-has-become-a-major-threat-to-food-security/?sh=5068ee3861c3}

\end{thebibliography}

\end{document}